\documentclass[12pt]{article}
\usepackage{setspace}
\usepackage{epsfig}
\usepackage{amssymb}
\usepackage{graphicx}
\def\be{\begin{equation}}
\def\ee{\end{equation}}
\def\ba{\begin{array}}
\def\ea{\end{array}}
\def\beqn{\begin{eqnarray}}
\def\eeqn{\end{eqnarray}}

\def\bt{\begin{tabular}}
\def\et{\end{tabular}}
\def\bc{\begin{center}}
\def\ec{\end{center}}

\begin{document}

\title{Comment on ``Texture Zeros and WB Transformations in the Quark Sector of the Standard Model"}

\author{Samandeep Sharma, Priyanka Fakay, Gulsheen Ahuja$^*$, Manmohan
Gupta\\ {\it Department of Physics, Centre of Advanced Study,
P.U.,
 Chandigarh, India.}\\
\\{\it $^*$gulsheen@pu.ac.in}}

 \maketitle

\begin{abstract}
Recently, using specific Weak Basis transformations, Y. Giraldo
[Phys. Rev. D 86, 093021 (2012)] has constructed some texture 5
and 4 zero quark mass matrices and examined their compatibility
with the quark mixing data. In this comment, we have re-analyzed
these to bring forth certain important issues regarding their
viability which need to be taken note of.

\end{abstract}

In the context of flavor physics, texture specific mass matrices
have provided valuable information to understand the flavor mixing
data \cite{ourrev}. The relationship of these mass matrices with
the Weak Basis (WB) transformations have been discussed by several
authors \cite{branco,costa,giraldo}. In particular, using WB
transformations, recently Y. Giraldo \cite{giraldo} has made an
attempt to explicitly construct texture five zero and texture four
zero quark mass matrices. While examining the compatibility of
these mass matrices with the mixing data, we come across certain
issues which need to be taken note of.

To begin with, we re-analyze a particular ``class" of texture 5
zero non-Fritzsch like quark mass matrices, discussed in section V
of \cite{giraldo} and shown to be compatible with the latest
mixing data, e.g.,
\be
 M_{U}=P^{\dagger}\left( \ba{ccc}
0 & 0 & |C_U|      \\ 0 & A_U &  |B_{U}|     \\
 |C_U| &     |B_{U}|  & \tilde{{B_{U}}} \ea \right)P, \qquad
M_{D}=\left( \ba{ccc} 0 & |C _{D}| & 0  \\
 |C_{D}| & 0 & |B_{D}|     \\
 0 &     |B_{D}|  &  A_{D} \ea \right).
\label{nft5}\ee This, however, appears to be somewhat in conflict
with the earlier analyses of similar matrices \cite{ourrev},
therefore making it necessary to re-examine the compatibility of
the above mentioned mass matrices with the recent data.

The essentials of the methodology usually used to carry out the
analysis include diagonalizing the mass matrices $M_U$ and $M_D$
by unitary transformations and obtaining a
Cabibbo-Kobayashi-Maskawa (CKM) matrix from these transformations.
To ensure the viability of the considered mass matrices, this CKM
matrix should be compatible with the quark mixing data, for
details regarding this we refer the readers to \cite{ourrev}.
Following this methodology for the above mentioned matrices
considered by Giraldo, the CKM matrix so obtained is given by
 \be V_{{\rm CKM}} = \left( \ba{ccc}
  0.9741-0.9744 &~~~~   0.2247-0.2260 &~~~~  0.0048-0.0119 \\
 0.0161-0.0833  &~~~~   0.0019-0.1461    &~~~~  0.0960-0.9890\\
0.0015-0.2288  &~~~~  0.0011-0.9755 &~~~~  0.0020-0.9999
\label{ckm} \ea \right). \label{3sm} \ee A look at this matrix
immediately reveals that the ranges of some of the CKM elements,
in particular of $|V_{ub}|$, $|V_{cd}|$, $|V_{cs}|$ and
$|V_{cb}|$, show no overlap with those obtained by recent global
analyses \cite{pdg}. This, therefore, leads one to conclude that
the texture 5 zero non-Fritzsch like quark mass matrices
considered in \cite{giraldo} are not compatible with the recent
quark mixing data.

\begin{figure}
\begin{center}
%\vspace{-0.5cm}
\includegraphics[width=70mm,angle=-90]{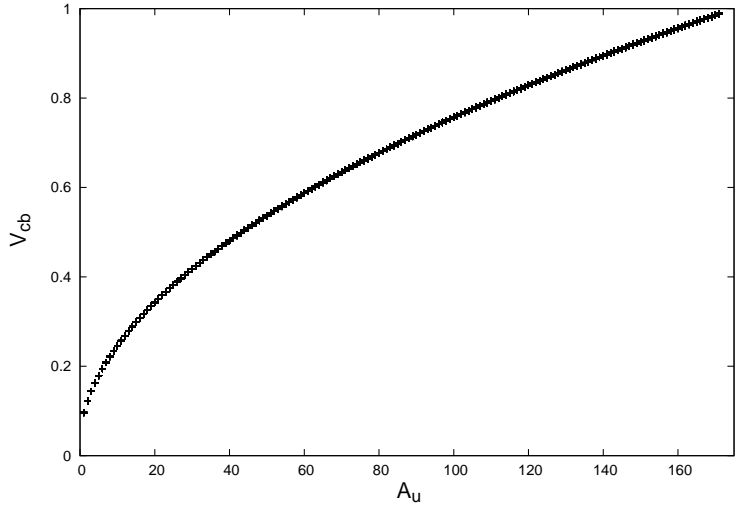}
 \end{center}
\caption{Plot showing the dependence of CKM matrix element
$V_{cb}$ on $A_U$, the free parameter of the mass matrix $M_{U}$.
The values of $A_U$ are in $GeV$ units.} \label{fig1}
\end{figure}

To re-emphasize our conclusion, we have examined the dependence of
one of the CKM matrix element $|V_{cb}|$, obtained from the mass
matrices considered here, solely on $A_U$, the free parameter of
the mass matrix $M_{U}$. This variation has been plotted in Figure
(\ref{fig1}) from which one can easily find that for all possible
values of $A_U$, the values of $|V_{cb}|$ are much larger than the
allowed range $0.0407-0.0423$  \cite{pdg}, clearly ruling out
these mass matrices.

The above conclusion is in direct conflict with that of Ref.
\cite{giraldo}, however it is not difficult to understand how a
fit for quark mixing matrix elements as well as the Jarlskog's
rephasing invariant parameter $J$ has been obtained in
\cite{giraldo}. It can be seen easily that this has been achieved
by introducing additional phases in the unitary matrices used for
diagonalizing the mass matrices $M_U$ and $M_D$ which cannot be
rephased away in the mixing matrix and hence can be considered
physical. To emphasize that these phases are physical or represent
additional parameters, we reconstruct the complex mixing matrix,
corresponding to the one given in equation (\ref{3sm}), e.g.,
\be
V_{{\rm CKM}} = \left( \ba{ccc}
  0.3447 + 0.911361 i &~~~~   0.02595 + 0.2233 i &~~~~  0.00893-0.00069 i \\
 -0.02391 -0.050292 i &~~~~   0.01658 - 0.01157 i   &~~~~  0.16240-6.4312 \times 10^{-5} i\\
0.00427 - 0.00755 i  &~~~~  -0.0230 + 0.0272 i &~~~~ 0.9932-0.0039
i \ea \right). \label{withph} \ee It may be mentioned that the
above matrix has been constructed using some typical values of
input parameters, e.g., \be A_U=4.0~GeV,\quad \phi_1=110^0,\quad
\phi_2=10^0, \ee which are within the ranges of these required to
reproduce the matrix in equation (\ref{3sm}).

Using the relation \cite{pdg}
\be
J\sum_{m,n=1}^{3}\epsilon_{ikm}\epsilon_{jln}={\rm Im}
(V_{ij}V_{kl}V_{il}^{*}V_{kj}^{*}), \label{j2} \ee  for the CKM
matrix in equation (\ref{withph}) the value of Jarlskog's
parameter $J$ is $5.564 \times 10^{-6}$. Even on allowing full
variation to all the input parameters, one gets $ |J| \le 5.6428
\times 10^{-6}$, clearly outside the range given by PDG 2012,
i.e., $(2.80-3.16) \times 10^{-5}$. This is in contrast with the
results given in \cite{giraldo}, therefore, the phases introduced
in are additional and physical ones and their addition leads to an
ad hoc fitting of the mixing matrix.

To further verify that additional phases introduced in
\cite{giraldo} are physical and cannot be rephased away, we have
investigated the implications of the mass matrices constructed in
\cite{giraldo} on other CP violating parameters as well. In this
context, we try to find the value of CP violating parameter
$\epsilon_k$ using the following rephasing invariant expression
given by Buras {\it et al.} {\cite{buras}}
 \be |\epsilon_K|
= \kappa_{\epsilon} \frac{G_F^2 F_K^2 m_K m_W^2}
        {6 \sqrt{2} \pi^2 \Delta m_k}B_K{\rm Im}\lambda_t
  [{\rm Re} \lambda_c(\eta_1 S_0(x_c)-\eta_3 S_0(x_c,x_t))
 - {\rm Re} \lambda_t \eta_2 S_0(x_t)], \label{eps} \ee
where $\eta_1$, $\eta_2$, $\eta_3$ are the perturbative QCD
corrections, $S_0(x_i)$ are Inami-Lim functions,
$x_i=m_i^2/M_W^2$, and $\lambda_i=V_{id}V^*_{is}$, $i=c,t$.
\begin{table}
\scriptsize

\bt{|c|c|p{1 cm} |} \hline & $ M_U
~~~~~~~~~~~~~~~~~~~~~~~~~~~~~~~~~~~~~~~~~~~~~~~~~~~~~~~~~~~~~~~~~~~M_D
$ & $\epsilon_k$ \\ \hline
 & & \\
  a& $\left ( \ba{ccc} {\bf 0} &{\bf 0} &
-92.3618+157.694i \\ {\bf 0}  & 5748.17 & 28555.1+5911.83i
\\-92.3618-157.694i & 28555.1-5911.83i & 166988 \ea \right )$,
 $\left ( \ba{ccc} {\bf 0} & 13.9899 & {\bf 0}
 \\ 13.9899  & {\bf 0} & 424.808 \\ {\bf 0}  & 424.808  &
2796.9 \ea \right )$ & $ 1.07038 \times 10^{-3}$ \\ & & \\

 b& $\left ( \ba{ccc} {\bf 0} & {\bf 0} & 123.038-285.496i \\
{\bf 0}  & 1430.03 & 18632.8-2336.25i \\ 123.038+285.496i &
18632.8+2336.25i  & 170033 \ea \right )$, $\left ( \ba{ccc} {\bf
0} & 13.2473 & {\bf 0} \\ 13.2473  & {\bf 0} & 425.817 \\ {\bf 0}
& 425.817  & 2796.6 \ea \right )$ & $ 0.90177
 \times 10^{-3}$ \\
 c & $\left ( \ba{ccc} {\bf 0} & 4543.2 & {\bf 0} \\ 4543.2 &
-171468 & 9388.13 \\ {\bf 0} & 9388.13 & 7.34102 \ea \right )$ ,
$\left ( \ba{ccc} {\bf 0} & 123.93+10.0184i & {\bf 0} \\
123.93-10.0184i & -2829.92 & 267.035+1.39152i \\ {\bf 0} &
267.035-1.39152i & 29.738 \ea \right )$ & $ 1.418 \times 10^{-3}$
\\ & & \\

\hline  \et \label{table1} \caption{Specific cases of texture 5
and 4 zero mass matrices constructed by \cite{giraldo} and their
corresponding $\epsilon_k$ values found here. }
 \end{table}

To this end, in Table 1 we present specific cases of texture 5 and
4 zero mass matrices considered by \cite{giraldo} and the
corresponding $\epsilon_k$ values found here. Interestingly, a
look at the table shows that these $\epsilon_k$ values are much
lower than its experimental value i.e. $(2.228\pm 0.011)\times
10^{-3}$ \cite{pdg}. This can be understood by examining the CKM
matrices corresponding to the numerical mass matrices presented in
Table 1. For example, for the first set of mass matrices presented
in row (a) of the table, we get the following numerical CKM matrix
  \be
 \left( \ba{ccc}
-0.484-0.845 \it{i} & 0.151+0.167 \it{i}  & -0.0028-0.002 \it{i}
\\ -0.22519-0.0012 \it{i} & -0.952 + 0.202 \it{i} &  0.029 + 0.029
\it{i}     \\ 0.0034 - 0.008 \it{i} & 0.0223 - 0.0336 \it{i} &
0.9991 - 0.0051 \it{i} \ea \right). \label{t51gckm} \ee For the
purpose of comparison, we have also constructed here the CKM
matrix using the latest values \cite{pdg} of Wolfenstein
parameters, e.g.,
 \be
 \left( \ba{ccc}
0.9746 & 0.2254  & 0.0012-0.0032 \it{i}     \\ -0.2254 & 0.9746 &
0.0412     \\
 0.0081-0.0032 \it{i} &  -0.0412 &  1 \ea \right).
\label{wckm}\ee A look at the above matrices clearly reveals that
most of the elements of mixing matrix given in equation
(\ref{t51gckm}) contain sizeable imaginary parts as compared to
those in equation (\ref{wckm}). In principle, one could say that
this may not have any implication in view of the facility of
rephasing invariance of CKM matrix. However, this is not the case
as can be found by subjecting the mixing matrix presented in
equation (\ref{t51gckm}) to CP violating parameter $\epsilon_k$.

We would also like to mention a couple of more points, e.g., for
the case of texture 4 zero quark mass matrices, well known to be
compatible with the mixing data \cite{ourrev}, while carrying out
the calculations pertaining to these in \cite{giraldo} again
arbitrary phase factors have been incorporated and some of the CP
violating parameters such as Sin2$\beta$ and $J$ have been
reproduced. Further, one finds that while constructing the non
parallel texture 4 zero quark mass matrices, \cite{giraldo}
assumes the (1,1)th entry to be zero because of large uncertainty
in it. This zero does not reflect a WB choice and thus the non
parallel texture 4 zero structure constructed by \cite{giraldo} is
effectively texture 3 zero only which re-emphasizes the conclusion
in \cite{branco} that starting with the most general quark mass
matrices it is not possible to obtain more than three texture
zeroes by any WB transformation. Further, it also needs to be
mentioned that while constructing parallel texture 4 zero
matrices, \cite{giraldo} does not start with the most general mass
matrices, rather it starts with a special weak basis wherein the
mass matrix $M_U$ has been taken to be diagonal and only the
matrix $M_D$ is considered to be most general.

To summarize, we have re-analyzed specific cases of texture 5 and
4 zero quark mass matrices considered in \cite{giraldo}. Texture 5
zero mass matrices have been shown to be incompatible with the
recent mixing data, in contrast with the findings of
\cite{giraldo} wherein additional phases have been incorporated
while showing the compatibility of these mass matrices. For some
of the cases of texture 5 and 4 zero mass matrices considered in
\cite{giraldo}, we find that even after incorporating additional
phases we are not able to reproduce the CP violating parameter
$\epsilon_k$. In conclusion, we would like to emphasize that one
needs to be careful in analyzing the implications of Weak Basis
transformations on textures.

  \vskip 0.5cm
{\bf Acknowledgements} \\G.A. would like to acknowledge DST,
Government of India (Grant No: SR/FTP/PS-017/2012) for financial
support. P.F., S.S., G.A. acknowledge the Chairperson, Department
of Physics, P.U., for providing facilities to work.


\begin{thebibliography}{99}

\bibitem{ourrev}M. Gupta, and G. Ahuja, Int. J. Mod. Phys. A {\bf 27},
1230033 (2012).

\bibitem{branco}G. C. Branco, D. Emmanuel-Costa and R. G. Felipe,
Phys. Lett. B {\bf 477}, 147 (2000).

\bibitem{costa} D. Emmanuel-Costa and C. Simoes,
Phys. Rev. D {\bf 79}, 073006 (2009).

\bibitem{giraldo}Y. Giraldo, Phys. Rev. D {\bf 86}, 093021 (2012).

\bibitem{pdg}J. Beringer {\it et al.}, Particle Data Droup,
Phys. Rev. D {\bf 86}, 010001 (2012).

\bibitem{buras}A. J. Buras, hep-ph/0101336.

\end{thebibliography}
\end{document}